\documentclass[aps,prx,twocolumn,letterpaper,superscriptaddress,showpacs]{revtex4-2}
\usepackage{graphicx}
\usepackage{CJK}
\usepackage{verbatim}
\usepackage{physics}
\usepackage[colorlinks,linkcolor=blue,anchorcolor=blue, citecolor=blue,urlcolor=blue,]{hyperref}
\usepackage{lineno}
\usepackage{bm}
\usepackage{mathptmx}
\usepackage{multirow}
\makeatletter
\newcommand{\parallelsum}{\mathbin{\!/\mkern-5mu/\!}}
\begin{document}
\begin{CJK*}{UTF8}{bsmi}
\title{Possible superconductivity in $d^{n} (n\neq 9)$ platforms ?}
\author{Zi-Jian Lang (\CJKfamily{gbsn}郎子健)}
\email{zjlang.wphy@gmail.com}
 
\date{\today}

\begin{abstract}
Superconductivity in square BX$_2$-plane-based materials  currently is only found in $d^9$ configuration.
This raises a question that whether other configurations $d^{n}$ ($n\neq 9$) of transition metals in the prototype of square BX$_2$ plane can also host superconductivity.
We systematically explore this question via analyzing the electronic structure of materials from $d^{8}$ to $d^5$ platforms using density-functional calculation, including existing materials SrFeO$_2$ and SrFeO$_2$F, and proposed ones LaNiO$_2$F and LaCoO$_2$F.
Results show a good commonality between these materials and the cuprate and nickelate superconductors in their high-energy electronic structure, namely dominant low-energy states by ligand $p$ and $d_{x^2-y^2}$ orbitals.
Other $d$ orbitals are all in the high-energy channel due to strong intra-atomic repulsion resulting in similar low-energy effective Hamiltonian except for the different number of local spins.
These results hopefully suggest the possible superconductivity besides the prototype of $d^9$.
Superconducting phases found in these sets of materials will be highly valuable to understand the high-temperature superconductivity and even to find better superconducting families than the cuprates.

\end{abstract}

\maketitle
\end{CJK*}

\section{Introduction}
High-temperature superconductivity (HTSC) in the layered cuprates~\cite{Muller} has attracted intensive studies after its discovery around thirty years ago~\cite{Dagotto,Damascelli}.
Although with great efforts, there is little progress on the improvement of transition temperature $T_c$ under ambient pressure after the highest $T_c\sim$130K found in Mercury-based cuprates~\cite{Schilling1993}.
This is basically due to the inadequacy in understanding the low-energy physics below 100 meV scale of superconductivity~\cite{Dagotto}.
The limitation of the choice of transition metal and ligand elements of the cuprates also restricts the parameter space to find better superconductors and the underlying physics.

To figure out the mechanism of unconventional superconductivity and also to search better superconductors with  higher $T_c$ (or even room-temperature ones), it is of significant importance to find the similarities and differences between different types of superconductors.
One good choice is iron pnictides and chalcogenides which have HTSC with $T_c\sim 50$K so far~\cite{Kamihara, Stewart2011}.
Different choices of ligand elements also provide rich tunability in the study of superconductivity~\cite{Stewart2011}.
However, all of these iron-based superconductors have completely different crystal structures.
The mutil-orbital nature in the low-energy physics makes the underlying mechanism much more complicated~\cite{Stewart2011} and perhaps distinguishable from the cuprates.

Fortunately, recently discovered nickelate superconductors~\cite{Li2019,Li2020} with similar layered structure and $d^9$ electronic structure of transition-metal atoms provide a good family of candidates to study the HTSC by the comparison with the cuprates.
Many properties, such as linear resistivity~\cite{Li2019}, domed-shaped superconducting phase~\cite{Li2020,Zeng2020}, and weakly insulating behaviors at low temperature~\cite{Li2019} all demonstrate similarities to the cuprates.
However, providing the weaker spin correlation~\cite{Lin2021} compared to that of the cuprates due to the large charge-transfer gap, it's questionable if one can find better superconducting properties in these NiO$_2$-based superconductors~\cite{Weber2012,Botana2020}.
Not to mention that the participation of Nd orbitals will further complicate ow-energy physics.

Thus, to find better superconductors, we need more families of superconductors as the cuprates analogs.
Despite the uncertain low-energy physics of HTSC in the cuprates, the most relevant orbitals for the superconductivity in high-energy scale are clear, which are just Cu-$d_{x^2-y^2}$ and 2 O-$p$ orbitals~\cite{Dagotto,Sawatzky1988}.
This significant simplification of Hilbert space in fact makes the cuprates an idea and clean system to study HTSC compared to other unconventional superconductors.
Therefore, it's highly valuable to explore possible superconductivity in other transition-metal-based layered materials with the same BX$_2$ plane.

However, to date, superconductivity is only found in square CuO$_2$-~\cite{Muller} and NiO$_2$-plane-based materials~\cite{Li2019} whose electronic configurations of transition metals in the parent compounds are both $d^9$.
Explorations of superconductivity in $d^8$ platform such as La$_2$NiO$_4$~\cite{Cava1991} and $d^7$ platform such as La$_2$CoO$_4$~\cite{Shinomori2002,Chichev2006} are all proven to be unsuccessful.
At low-doping level, doped carriers are strongly localized instead of moving ``freely'' in the cuprates.
This is probably due to the strong coupling to phonons or magnons~\cite{Anisimov1992,Bi1993,Zaanen1994,Dobry1994}.
Is that strong coupling simply caused by the large spins of transition metals or the special crystal structures of these materials?
It is still unclear whether other platforms of $d^n$ (where $n=5,6,7,8$) could support the existence of superconductivity if doped holes can move.
Therefore, it is interesting to investigate the behaviors of doped holes in other types of materials with the square BX$_2$ plane.

Here, to explore the possible superconductivity like the cuprates, we investigate the electronic structure of materials with the same square BX$_2$ plane via density functional calculation.
These include different electronic configurations of $d^8, d^7, d^6$, and $d^5$ platforms.
We propose the several candidates to host these states include existing materials SeFeO$_2$($d^6$), and SeFeO$_2$F($d^5$), and designed materials LaNiO$_2$F($d^8$), LaCoO$_2$F($d^7$).
The resulting electronic structures of these materials show good similarities with the cuprates and the nickelates in the high-energy scale.
Other orbitals than $d_{x^2-y^2}$ in these materials are all in high energy suggesting the irrelevance of them to the low-energy physics of hole carriers.
These demonstrate the high possibility of superconductivity in hole-doped samples.
Comparison between these families of materials and the cuprates and previous iron-based superconductors should give useful insight to understand HTSC.
The rich choice of transition metals and ligands also pave the way to find higher-$T_c$ superconductors than the cuprates in the future.

The contents of this paper are listed as follows.
In Section~\ref{choice_t_l}, we discuss the basic rules in the choice of transition metals and ligands given a large number of different elements.
Section~\ref{results} shows the resulting electronic structure for different materials with platforms from $d^8$ to $d^5$.
In Section~\ref{discussion}, we show the general Hamiltonians for these materials and the comparison between them.
The remaining open questions are also discussed.
Last, we give the summary of our results in Section~\ref{summary}.

\section{Choice of transition metals and ligands}
\label{choice_t_l}

Let's first discuss the basic conditions to find proper materials with square BX$_2$ plane.
In fact, as the cuprates analogs, the necessary conditions should be the single occupation of $d_{x^2-y^2}$ orbitals($d_{x^2-y^2}^1$), and the full occupation of ligand $p$ orbitals ($p^6$).
This suggests that transition metals in groups 7-10 all could be the possible choice.
From the high-energy consideration, as long as the irrelevant $d$ orbitals are far away from the Fermi surface such that they are all in high energy, the occupation number of these orbitals should not influence the low-energy physics significantly of carriers living in $d_{x^2-y^2}$ and ligand $p$ orbitals.
This indicates that other $d^{n}$ ($n\neq9$) platforms can probably allow the superconductivity like the cuprates.
Since high-spin states in $4,5$-$d$ metals are less common given the smaller Hund's coupling than the crystal splitting under BX$_2$ environments, $d_{x^2-y^2}$ orbital is often empty in $4,5d$ transition-metal ions.
Thus, we mainly focus on the 3$d$ metal, Mn, Fe, Co, Ni for simplicity.

Concerning the choice of ligands, elements in groups 15-17 should all be the possible candidates.
However, the ionic radius of $3,4,5$-$p$ ions are usually large such that square BX$_2$ plane is hard to realize for these elements.
Therefore, we will mainly focus on the $2$-$p$ atoms.
Given the high-energy $p$ orbitals and high valence of N$^{3-}$, it's not easy to find the corresponding transition-metal ions to ensure the $p^6$ state.
The BN$_2$ type of materials will be ignored in our discussions.
F$^-$ has extremely low-energy $p$ orbitals, which leads to a very large CT gap for most transition metals.
Thus, it is also not a good choice.
Therefore, O$^{2-}$ is the remaining choice which allows the existence of BO$_2^{2-}$ for divalent ions and BO$_2^{-}$ for trivalent ions.
We next will focus on the BO$_2$-type of materials from $d^5$ to $d^8$ platforms.

\begin{figure*}
	\begin{center}
		\includegraphics[width=12cm]{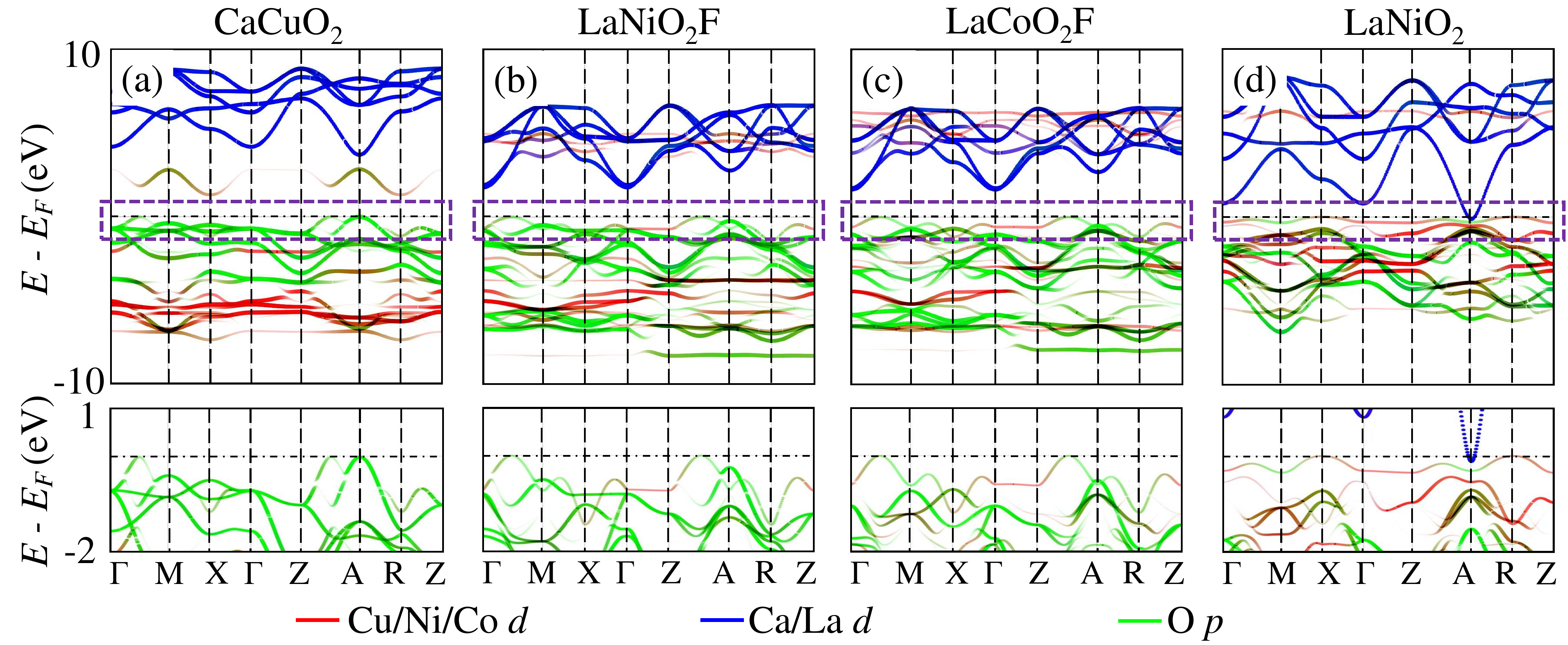}
	\caption{Comparison of the electronic structure between (a) CaCuO$_2$, (b) LaNiO$_2$F, (c) LaCoO$_2$F, and (d) LaNiO$_2$ all under antiferromagnetic ordered phase. The red, blue and green colors denote the weight of transition metal $d$ orbitals, Ca/La $d$ orbitals, and O $p$ orbitals respectively. For a clear view, we unfold them into the Brillouin zone corresponding to one-Cu/Ni/Co unit cell.
	The lower panels show the zoom-in band structure denoted by purple boxes in the upper panels. }
	\label{fig1}
	\end{center}
\end{figure*}

\section{Results}
\label{results}
\subsection{$d^8$ platform}
\label{d8}
The possible ions to realize this $d^8$ configuration could be Cu$^{3+}$, and Ni$^{2+}$.
For Cu, due to the low energy of $d$ orbitals, the energy of state $d^9\underline{L}$ is lower than that of $d^8L$, such that this $d^8$ configuration is hard to realize in CuO$_2$-based materials.
Therefore, Ni$^{2+}$ is the remaining candidate for this type of platform.
To ensure the single occupation of $d_{x^2-y^2}$ orbital, we need to remove one additional electron from another $d$ orbital.
The simplest choice is to add a ligand atom in $z$ direction.
There is a report of $d^8$ state of Ni$^{2+}$ in A$_2$NiO$_2$X$_2$ (X could be halogens or hydrogen) but with low-spin state~\cite{Kitamine2020} suggesting the possible superconductivity via electron doping.
However, the introduction of apical atoms will inevitably increase the energy of $d_{3z^2-r^2}$ orbital resulting in the single occupation of two $e_g$ orbitals under the presence of strong local repulsion within $d$ orbitals.
To avoid the possible influence of $d_{3z^2-r^2}$ and apical ligand orbitals~\cite{Takeda1990,Cava1991}, fluorine with much lower-energy $p$ orbitals than that of O is a good candidate.
Given the similar ionic radius of F$^-$ and O$^{2-}$, the intercalation of F in the nickelates should be easily achieved.
Here, we mainly focus on the fluorinated LaNiO$_2$, namely LaNiO$_2$F where the position of F is (0,0,0.5) in reduced coordinate.

We next calculate the electronic structure of this LaNiO$_2$F.
The lattice constants are assumed to be $a=b=3.95\text{\AA},c=4.0\text{\AA}$ since F$^-$ is slightly larger than O$^{2-}$.
(These values are borrowed from SrFeO$_2$F since electronic structure should be insensitive to the small change of lattice constants if the system is not close to phase transitions.)
Figure~\ref{fig1} shows the resulting unfolded electronic structure~\cite{Wei2002,Wei2010} of LaNiO$_2$F under AFM ordered phase.
The system is an insulator with $\sim2$eV band gap.
Compared to LaNiO$_2$ and CaCuO$_2$, the energy of $d$ orbitals is lower than that of LaNiO$_2$ reflecting the more stable nature of Ni$^{2+}$ state than that of Ni$^+$, and higher than Cu$^{2+}$ due to their chemical nature.
There are two upper Hubbard bands from two Ni $e_g$ orbitals corresponding to the high-spin state of Ni$^{2+}$.

Here, the coupling between $d_{x^2-y^2}$ and O $p$ orbital is $t_{pd}\approx1.3$~eV similar to the cuprates and nickelates.
Assuming charge-transfer gap $\Delta_{CT}=5$~eV, $U_{dd}=8$~eV, $U_{pp}=4$~eV, the resulting energy gaining from the exchange process for the intrinsic holes in $e_g$ orbitals is $\Delta E=2t_{pd}^4(\frac{1}{\Delta_{CT}^2U_{dd}}+\frac{2}{\Delta_{CT}^2(2\Delta_{CT}+U_{pp})})\approx0.061$~eV which is similar to the cuprates but apparently larger than that of the nickelates.
(For easy comparison with the cuprates and nickelates with only one local spin, we do not use the exchange parameter here but advocate the energy scale of magnetic correlation. 
The energy difference of the same spin and opposite spin in two nearest neighboring sites is $\frac{1}{2}J\approx0.075$~eV for the cuprates~\cite{Ogata2008} which is also the energy gain from superexchange processes.)

As similar to CaCuO$_2$, the states near the fermi energy are mainly contributed from O $p$ orbitals.
The energy of apical F $p$ orbital is much lower than that of O $p$ and Ni $d$ orbitals as expected, and thus is unrelated to the low-energy physics of hole carriers if the system is hole-doped.
This is significantly different from other nickelates with Ni$^{2+}$, such as La$_2$NiO$_4$.
Since even under very large charge-transfer gap in NdNiO$_2$~\cite{Botana2020,Zhang2020}, doped holes still reside in O $p_{\parallelsum}$ and Ni $d_{x^2-y^2}$ orbitals to form Zhang-Rice singlet~\cite{Zhang} like the cuprates.
($p_{\parallelsum}$ means the orbitals that point from ligand to transition metal.)
Additional holes should still reside in O $p$ orbital in LaNiO$_2$F though $d_{3z^2-r^2}$ now is singly occupied.
Given that $d_{3z^2-r^2}$ is far away from the Fermi energy, the presence of one local spin in this orbital can only influence carriers in $d_{x^2-y^2}$ and O $p$ orbital via spin correlation.
Whether this coupling can influence the movement of hole carriers or further possible superconductivity needs more investigation.

\subsection{$d^7$ platform}
\label{d7}

Now let's consider the $d^7$ platform.
Ni$^{3+}$ and Co$^{2+}$ are both possible ions to realize this $d^7$ electronic configuration.
To keep the stability of square BX$_2$ plane, the occupation number of  $d_{xz}, d_{yz}$ orbitals needs to be the same.
Otherwise, the unbalanced occupation in these two orbitals will lead to additional symmetry breaking in $x,y$ direction, such as orbital order~\cite{Lee2009}.
To avoid this effect, the possible electronic high-spin (HS) state is $d_{x^2-y^2}^1d_{3z^2-r^2}^1d_{xy}^1d_{xz}^2d_{yz}^2$, and $d_{x^2-y^2}^1d_{3z^2-r^2}^2d_{xy}^2d_{xz}^1d_{yz}^1$.
From the crystal-field consideration, the former is allowed in $z$-direction elongated octahedral environment while the latter is not realized in the presence of BX$_2$ plane.
Another possible configuration is low-spin(LS) state of $d_{x^2-y^2}^1d_{3z^2-r^2}^0t_{2g}^6$. 
To achieve this state, one needs to add high pressure along the $z$ direction (or strain in $xy$ plane) to push the energy of $d_{3z^2-r^2}$ orbital higher than that of $d_{x^2-y^2}$.
In this case, the system still has 1/2 local spin in $d_{x^2-y^2}$ orbital like the $d^9$ platform in the cuprates and nickelates but with a fully empty $d_{3z^2-r^2}$ orbital.

{\it High-spin $d^7$ platform -- Co$^{2+}$.}
Let's first investigate the possible electronic structure under the presence of HS-Co$^{2+}$.
To keep the correct valence of Co, the perovskite structure such as LaCoO$_2$X (where X could be halogens), and Ruddlesden-Popper-type structure such as La$_2$CoO$_4$ are both allowed.
As similar to the case of Ni$^{2+}$, to avoid the possible involvement in the low-energy physics of apical oxygen, we substitute oxygen with fluorine.
For simplicity, we take LaCoO$_2$F as an example.
Here, the coordinate of F is (0,0,0.5) and the corresponding lattice constants are the same as LaNiO$_2$F.

Figure~\ref{fig1}(c) shows the resulting electronic structure of HS-LaCoO$_2$F under AFM ordered phase.
There are three upper Hubbard bands corresponding to the single occupation of $d_{x^2-y^2}, d_{3z^2-r^2},d_{xy}$ orbitals as expected.
Under the same +2 valence, one can clearly observe the increase of the energy of $d$ orbitals from Cu$^{2+}$ to Ni$^{2+}$ and Co$^{2+}$ which is consistent with chemical intuition for different transition metals.
This can lead to the increase of CT gap and thus the reduction of CT characteristics.
Nevertheless, compared to LaNiO$_2$, the weight of the O $p$ orbitals near the Fermi energy is still considerably larger.
The energy of F $p$ orbitals again is extremely low, and thus we can safely drop the F $p$ and Co $d_{3z^2-r^2}$ orbitals.

Here, the coupling between O $p$ and Co $d_{x^2-y^2}$ orbitals is $t_{pd}=1.40$~eV, slightly larger than that of the cuprates and nickelates.
Given the similar bond length in our calculations, this stronger coupling probably is due to the smaller number of electrons in Co$^{2+}$ compared to Cu$^{2+}$ and Ni$^{2+}$.
More importantly, a stronger $pd$ coupling should support better magnetic correlation in this material.
Assuming the parameters $\Delta_{CT}=5.5$~eV, $U_{dd}=8$~eV, $U_{pp}=4$~eV, we can obtain the energy gain from superexchange processes $\Delta E=0.066$~eV which is much larger than that of LaNiO$_2$ as a result of smaller charge-transfer gap and larger $pd$ coupling.

{\it High-spin $d^7$ platform -- Ni$^{3+}$}.
Besides the high-spin Co$^{2+}$ ion, high-spin Ni$^{3+}$ is also the possible candidate.
However, in the elongated octahedral environment, the doped holes in fact reside in O $p$ orbital and Ni $d_{x^2-y^2}$ from the results of LaNiO$_2$F.
This means if we substitute La with alkaline-earth metal, we can only obtain the low-spin state of Ni$^{3+}$ with empty $d_{x^2-y^2}$ orbital.
Thus, whether the high-spin Ni$^{3+}$ can be stabilized in octahedral environment to keep the single occupation of $d_{x^2-y^2}$ is questionable.
The possible option to increase the energy of $d_{xy}$ orbital such that $d^{7}L$ is more stable than $d^8\underline{L}$.
Of course, which kind of crystal field would allow this particular energy splitting is out of the scope of this paper. 
We'll not discuss this high-spin Ni$^{3+}$ in detail here.

\begin{figure*}
	\begin{center}
		\includegraphics[width=12cm]{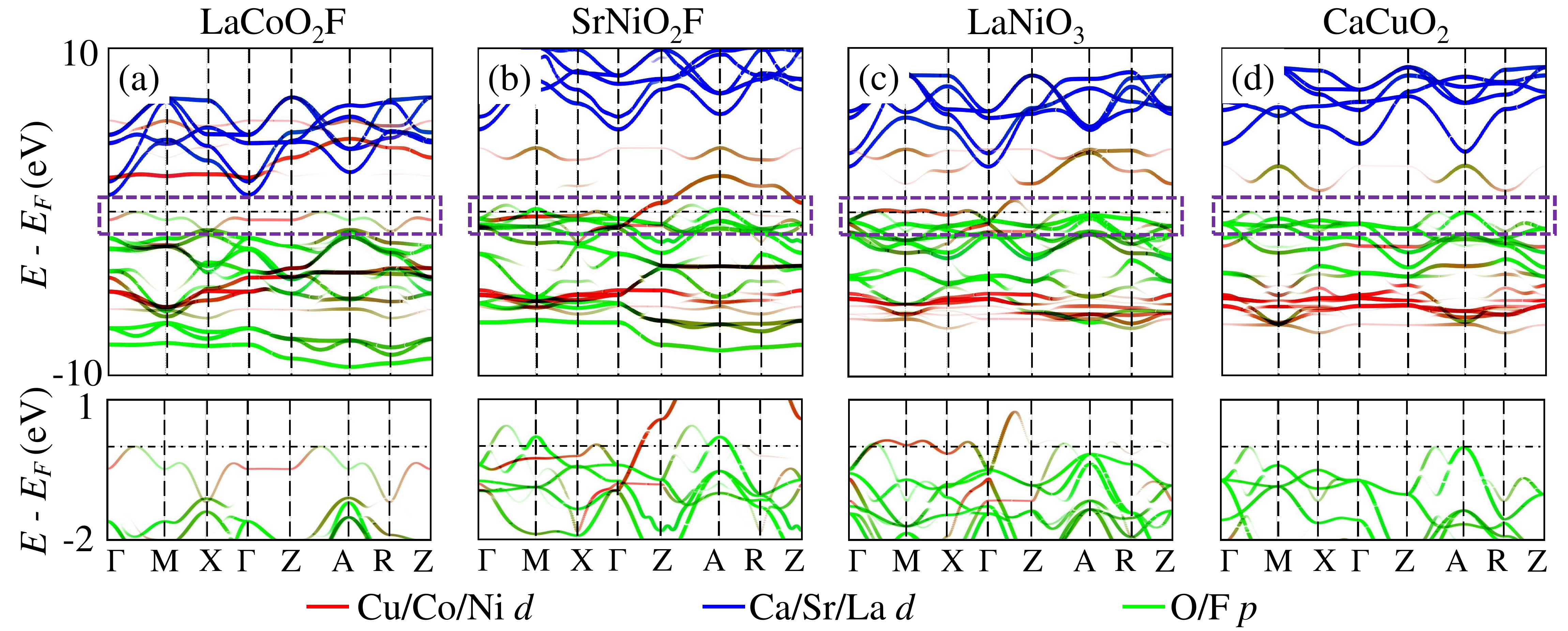}
	\caption{Comparison of unfolded electronic structure between (a) LaCoO$_2$F, (b)SrFeO$_2$F, (c) LaNiO$_3$, and (d)CaCuO$_2$. The lattice structure of LaCoO$_2$F, SrFeO$_2$F, and LaNiO$_3$ are assumed to be compressed in $z$ direction to realize low-spin $d_{x^2-y^2}^1t_{2g}^6$ state.
	The red, blue and green colors represent the weight of transition metal $d$, Ca/Sr/La $d$, and O/F $p$ orbitals respectively. The lower panels show the zoom-in figures denoted by purple boxes in the upper panels.}
	\label{LSd7}
	\end{center}
\end{figure*}

{\it Low-spin $d^7$ platform.}
Besides the HS $d^7$ state, the LS $d^7$ shows an interesting electronic structure much close to $d^9$ state.
The $d_{x^2-y^2}^1d_{3z^2-r^2}^0t_{2g}^6$ state has the same localized spin in $d_{x^2-y^2}$ orbital like the cuprates.
Unlike other $d^n$ platforms, there are even no additional spins in $t_{2g}$ and $d_{3z^2-r^2}$ orbitals to influence the low-energy carriers via spin correlation.
The only difference compared to the $d^9$ platform is the existence of empty $d_{3z^2-r^2}$ orbital.
Nevertheless, the $d_{3z^2-r^2}$ orbital is in high energy and can be integrated out resulting in the almost same low-energy physics as similar to the cuprate superconductors.
Thus, we expect this type of configuration should be highly possible to host superconductivity.

To realize this low-spin configuration of $d^7$ platform, elements in group 9 with +2 valence and elements in group 10 with +3 valence are all possible.
For +2 valence, transition metals usually are stable under this valence.
For +3 valence, unlike Cu$^{3+}$ ion, the energy of $d$ orbitals of group-10 atoms such as Ni, are higher, such that $d^7L$ should be allowed in the parent compounds~\cite{Upton2015}.
In fact, previous studies~\cite{Chaloupka2008,Hansmann2009,Disa2015} already proposed LS Ni$^{3+}$ state as the candidate to host superconductivity.
Therefore, let's first assume the LS $d^7$ state for group 9\&10 transition metals is realizable.

For simplicity, here we mainly focus on the 3$d$ transition-metal ions, namely Co$^{2+}$ and Ni$^{3+}$.
For Co$^{2+}$ with stable $d^7$ ground state, adding pressure in $z$ direction (or strain in $xy$ plane) of LaCoO$_2$F and SrNiO$_2$F or existing ReNiO$_3$(Re is the rare-earth element) should allow this state.
The pressure along $c$ axis would break the symmetry of the octahedral field to realize strong anti-Jahn-Teller effect such that Hund's coupling can be overcome.
In this case, the only electron in $e_g$ orbitals would be in $d_{x^2-y^2}$ orbital.

To investigate the possibility of LS-$d^7$ state, we take compressed LaCoO$_2$F, SrNiO$_2$F, and LaNiO$_3$ as examples, and then study their electronic structure.
The lattice constants are set as $a=b=3.95~\text{\AA},c=3.7~\text{\AA}$ for all materials.
Here, we set a smaller $c$ in order to increase the energy of $d_{3z^2-r^2}$ orbital.
All of them are set in P4/mmm space group.

Figure~\ref{LSd7} shows the resulting electronic structure of LaCoO$_2$F, SrNiO$_2$F, and LaNiO$_3$, all under AFM order.
As we expect, LaCoO$_2$F is an insulator with $\sim1$~eV band gap where $d_{3z^2-r^2}$ orbital is fully empty and $d_{x^2-y^2}$ orbital is singly occupied.
Near the fermi surface, the low-energy states are dominated by O $p$ and Co $d_{x^2-y^2}$ orbitals.
One can thus integrate any other irrelevant orbital to obtain the low-energy Hamiltonian with only Co $d_{x^2-y^2}$ and 2 O $p_{\parallelsum}$ orbitals.
In this case, $t_{pd}=1.40$~eV, thus the energy gain from superexchange process is $\Delta E=0.066$~eV where we use $\Delta_{CT}=5.5$~eV, $U_{dd}=8$~eV, $U_{pp}=4$~eV.
Under such magnetic correlation, hole-doped LS-LaCoO$_2$F should be a better superconductor than the infinite-layer nickelates.

Note that this LS Co$^{2+}$ state is influenced by the choice of parameter $U$ which is however fixed in real materials.
For $3d$ transition metal with strong Hund's coupling, whether the crystal-field splitting is large enough to overwhelm the energy gain from Hund's coupling still needs experimental investigations.
From practical consideration, if we are not limited in $3d$ atoms, $4d/5d$ transition metals, which usually prefer LS configuration, are probably the better candidates to realize this specific electronic state.

Now, let's consider another candidate LS-Ni$^{3+}$.
For LS Ni$^{3+}$, the charge-transfer characteristic is strongly enhanced compared to Co$^{2+}$ due to the low energy of $d$ orbitals in Ni$^{3+}$[cf. Fig.~\ref{LSd7}(b)(c)] as expected.
The CT gap between green bands from O $p$ orbitals and the red upper band from Ni $d_{x^2-y^2}$ orbitals is also similar to that of CaCuO$_2$.
These seem to suggest good superconductivity in the system.

However, as shown in Fig.~\ref{LSd7}(b)\&(c), there is always one band from Ni $d_{3z^3-r^2}$ orbital crossing the fermi energy (even if we force the system to start from LS-$d^7$ Ni$^{3+}$ state.)
This implies that the $d_{3z^3-r^2}$ is not fully empty.
These additional electrons in fact originate from O $p$ orbitals which indicates a $d^8\underline{L}$ state.
This is not surprising since materials with Ni$^{2+}$ are already charge-transfer insulators as shown in Section~\ref{d8}.
Thus, starting from $d^8$ state, doped holes should reside in O $p$ orbitals due to the fact that compression in $z$ direction is not strong enough to decrease the energy $d_{3z^2-r^2}$.
Thus, whether this $d_{x^2-y^2}^1d_{3z^2-r^2}^0t_{2g}^6$ can be established in Ni$^{3+}$ probably still needs more studies in the future.
(See Appendix~\ref{app2} for a detailed discussion on this issue.)

\begin{figure*}
	\begin{center}
		\includegraphics[width=12cm]{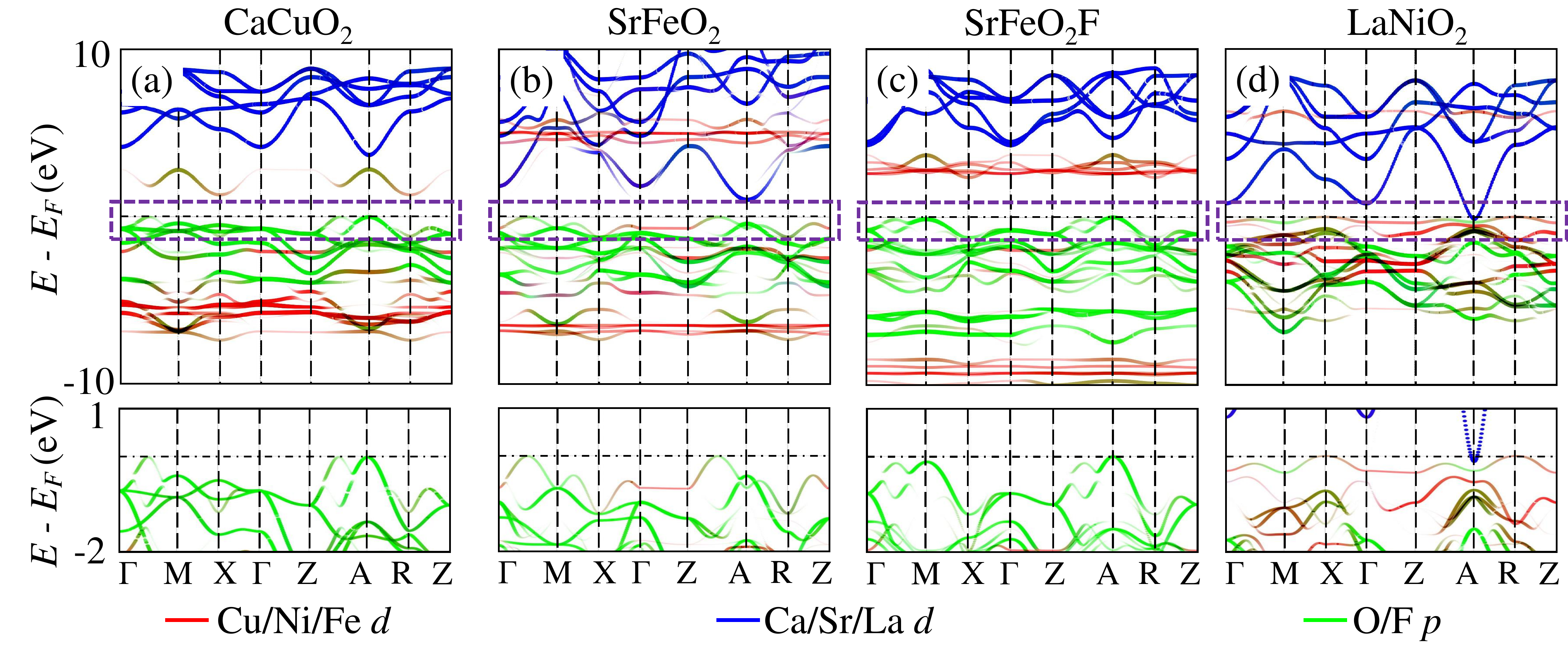}
	\caption{Comparison of unfolded electronic structure between (a)CaCuO$_2$, (b)SrFeO$_2$, and (c) LaNiO$_2$. The red, blue and green colors represent the weight of transition metal $d$, Ca/Sr/La $d$, and O $p$ orbitals respectively. The lower panels show the zoom-in figures denoted by purple boxes in the upper panels.}
	\label{d6d5bs}
	\end{center}
\end{figure*}

\subsection{$d^6$ platform}
\label{d6}
The possible transition metals to achieve $d^6$ configuration are basically Fe$^{2+}$ and Co$^{3+}$.
In fact, the square FeX$_2$ plane has already been  realized in experiments.
SrFeO$_2$~\cite{Tsujimoto2007}, a stable infinite-layer compound, has a similar structure to CaCuO$2$ and LaNiO$2$.
Furthermore, the parent compound is also an antiferromagnetically ordered insulator~\cite{Tsujimoto2007} below N\'eel temperature $T_N=473$K with $S$=2 local moment different from the spin $1/2$ local moment in the cuprates and nickelates.
Thus, it's very interesting to study the properties of carrier-doped materials.
However, there are only Eu substitution~\cite{Matsuyama2010} which shows much reduced resistivity compared to the parent compound, and Sm substitution resulting in good metallic behavior~\cite{Katayama2019}.
Both of them can only add electrons to the system.
Currently, no experiment explores the hole-doped samples~\cite{Tassel2012}.
Therefore, investigations on the behaviors of doped holes in these materials are necessary.

To investigate the low-energy physics of SrFeO$_2$ of hole carriers, let's first focus on its high-energy electronic structure.
The lattice constants used here are $a=b=3.99\text{\AA},c=3.47\text{\AA}$~\cite{Tsujimoto2007} in our calculation.
Figure~\ref{d6d5bs} shows the resulting band structure of SrFeO$_2$, CaCuO$_2$ and LaNiO$_2$ in 20-eV scale from density functional calculation using LDA+$U$ approximation~\cite{Anisimov,wien2k} for the AFM ordered phase.
Obviously, SrFeO$_2$ is an AFM insulators which is consistent with experimental results~\cite{Tsujimoto2007}.
Fig.~\ref{d6d5bs}(b) shows that there are four unoccupied $d$ orbitals(in red) in the spin minority channel of iron corresponding to the high-spin $d^6$ configuration of Fe$^{2+}$, and the only double occupied orbital is the $d_{3z^2-r^2}$ orbital as reported before~\cite{Xiang2008}.
As similar to CaCuO$_2$, near the fermi surface, there are three relevant orbitals --- 2 O-$p$ orbitals(in green), and Fe $d_{x^2-y^2}$ orbital.
Note that, as opposed to the cuprates, doped electrons will mainly reside in Sr $d$ orbitals as shown in Fig.~\ref{d6d5bs}(b) instead of $d_{x^2-y^2}$ orbital of transition metal.
Again, using the same parameters of $\Delta_{CT}=5.5$~eV $U_{dd}=8$~eV, and $t_{pd}=1.45$~eV, the energy gain from exchange process is $\Delta E=0.075$~eV.

Since the energy of Fe orbitals is higher than that of Ni in chemical nature, the weight of the low-energy bands are expected to be dominated by the Fe-$d$ orbitals like LaNiO$_2$[c.f Fig.~\ref{d6d5bs}(c)]~\cite{Lee,Botana2020,Karp2020, Lang2020}.
However, as shown in Figure~\ref{d6d5bs}(b), the O-$p$ orbitals (in green) instead dominate the low-energy states of SrFeO$_2$ below Fermi surface which is more closed to CaCuO$_2$ in Fig.~\ref{d6d5bs}(a).
This in fact reflects the unstable nature of Ni$^{1+}$ ion compared to Fe$^{2+}$.
On the other hand, these much lower-energy $d$ orbitals of Fe$^{2+}$ in turn simplify the low-energy physics of hole carriers in SrFeO$_2$.
Though there are three additional holes in the Fe $t_{2g}$ orbitals, all of them are high-energy orbitals and decoupled from the relevant $d_{x^2-y^2}$ and O-$p$ orbitals except for the spin channels.
One can thus safely integrate out charge degree of freedom of these $t_{2g}$ orbitals to obtain a similar low-energy physics to the cuprates.

{\it Co-based type.}
Besides the Fe$^{2+}$, Co$^{3+}$ with the same $d^6$ configuration is also the possible candidate.
Especially, Co$^{3+}$ has even lower-energy $d$ orbitals than that of Cu$^{2+}$ indicating an ideal platform to host strong CT nature and thus possibly better superconducting properties.
To ensure the single occupation of $d_{x^2-y^2}$ orbital and to avoid the possible symmetry breaking along $x,y$ direction, the electronic configuration could be $d_{x^2-y^2}^1d_{3z^2-r^2}^2d_{xy}^1d_{xz}^1d_{yz}^1$ identical to that of Fe$^{2+}$ in SrFeO$_2$, and $d_{x^2-y^2}^1d_{3z^2-r^2}^2t_{2g}^4$ in some pyramidal crystal field~\cite{Knee2003}.
However, for the former, there is no known ACoO$_3$ (A is +1 ion) with perovskite structure which usually is the precursor to synthesis ACoO$_2$.
And for the latter, to avoid the influence of apical ligand $p$ orbitals, the apical atoms should be F and distribute orderly which is also not realized in experiments.
Therefore, we will not discuss this electronic configuration in detail here.
(See Appendix for more details about the electronic structure of materials with HS-Co$^{3+}$.)

\subsection{$d^5$ platform}
\label{d5}
The possible ions to allow $d^5$ configuration are Fe$^{3+}$ and Mn$^{2+}$.
To ensure the single occupation of $d_{x^2-y^2}$ orbital, the electronic configuration could be a simple HS-$d^5$ state in which all $d$ orbitals have one electron.
$d_{x^2-y^2}^1t_{2g}^4$ (such as $d_{x^2-y^2}^1d_{xy}^2d_{xz}^1d_{yz}^1$) is also allowed in some anti-Jahn-Teller distorted environment.
Here for simplicity, let's focus on the HS-$d^5$ state with $S=5/2$ local spins.

{\it Fe-based type.}
For Fe$^{3+}$, to achieve this electronic state, we can use AFeO$_2$ (A is the alkali metal) as the prototype.
However, given the results in the previous section that doped holes mostly reside in O $p$ orbitals.
This type of material with one hole in O $p$ orbitals is probably not the applicable choice.
The remaining available approach is to increase the energy of $d_{3z^2-r^2}$ such that both $e_g$ orbitals are singly occupied.
Therefore, we can add halogen as the ligand in the $z$ direction.

As similar to the fluorination of the nickelates, this $d^5$ configuration can be realized by intercalation of fluorine in SrFeO$_2$.
Previous studies already demonstrate the high-spin state of Fe$^{3+}$ in SrFeO$_2$F where all $d$ orbitals are singly occupied.
Specially, N\'eel temperature increases to $T_N\sim$700K~\cite{Corey2014,Colin2014} compared to SrFeO$_2$.
Note that fluorination via XeF$_2$ seems to lead to the random distribution of F/O although starting with SrFeO$_2$~\cite{Colin2014}.
In this case, there is no homogeneous BX$_2$ plane anymore and the $d_{x^2-y^2}$ and $d_{3z^2-r^2}$ will both involve in the low-energy physics.
Fortunately, fluorination via polyvinylidene fluoride from SrFeO$_2$ may generate the ordered distribution of F/O given the resulting tetragonal structure~\cite{Katayama2014}.

Assuming the apical F can be realized in experiments, we next calculate the electronic structure of SrFeO$_2$F to investigate the influence of apical fluorine.
The lattice constants are set as $a=b=3.95\text{\AA},c=4.0\text{\AA}$~\cite{Katayama2014}.
Figure~\ref{d6d5bs} shows the band structure of SrFeO$_2$F and its comparison to CaCuO$_2$, SrFeO$_2$, and LaNiO$_2$, all under the same AFM ordered phase.
Obviously, SrFeO$_2$F is also an insulator with $\sim2.5$~eV charge-transfer gap between Fe $d$ orbitals and O $p$ orbitals.
There are 5 upper Hubbard bands above the chemical potential corresponding to the high-spin $d^5$ state of Fe$^{3+}$ which is consistent with the experimental observations.

From SrFeO$_2$ to CaCuO$_2$, the energy of upper Hubbard bands gradually decreases as a result of reducing energies of $d$ orbitals.
In fact, if we compare the energy of the lower Hubbard band, Fe$^{3+}$ $d$ orbitals is apparently lower than that of Cu$^{2+}$.
The large energy splitting between LHB and UHB in SrFeO$_2$F is due to Hund's coupling in the presence of 5 local spins in Fe$^{3+}$.
Nevertheless, compared to SrFeO$_2$, the CT gap still reduces thanks to the low-energy $d$ orbital in Fe$^{3+}$.
As similar to LaNiO$_2$F and LaCoO$_2$F, the energy of F $p$ orbital is about $-9$~eV, far away from the Fermi energy.
The electronic structure near the Fermi energy shows good similarity to CaCuO$_2$.
Low-energy states all originate from O $p$ orbitals.
One thus expects that doped holes will mostly reside in O $p_{\parallelsum}$ orbitals to form similar singlet states with intrinsic holes in $d_{x^2-y^2}$.

The resulting coupling between O $p$ and Fe $d_{x^2-y^2}$ increases to $t_{pd}=1.50$~eV given the small occupation number of Fe $d^6$ orbitals.
In this case, with the same $U_{dd}=8$~eV, $U_{pp}=4$~eV, and estimated charge-transfer gap $\Delta_{CT}\approx$5eV, the energy gain from superexchange process is $\Delta E=0.11$eV which is apparently larger than that of the cuprates due to larger $pd$ coupling.
Compared to SrFeO$_2$, this enhancement of $\Delta E$ mainly originates from the reduction of CT gap as a result of lower-energy $d$ orbitals.
This strong magnetic correlation is also consistent with the high N\'eel temperature observed in experiments~\cite{Corey2014}.

{\it Mn-based type.} 
In addition to Fe$^{3+}$, Mn$^{2+}$ is also the possible ion to host HS-$d^5$ state.
To ensure the presence of MnO$_2$ plane, one simple choice is to add apical fluorine atoms like that in SrFeO$_2$F.
Therefore, LaMnO$_2$F could be the candidate which can be synthesized via topotactical fluorination of LaMnO$_2$ or LaMnO$_3$.
However, LaMnO$_2$ has not been realized in experiments currently.
For LaMnO$_3$, how to ensure the correct substitution of O with F is also a problem.
In addition, Mn$^{2+}$ with high-energy $d$ orbitals can probably only support a weak magnetic correlation.
Thus, we will not discuss Mn$^{2+}$ in detail here.

\section{Discussion}
\label{discussion}
From the electronic structure of these different platforms, we can thus write a generic low-energy Hamiltonian of hole carriers for all these materials like the two-band model in the cuprates~\cite{Emery1987,Emery1988,Zaanen1988} as:
\begin{equation}
\begin{split}
    H &=\sum_{ijmn\mu}t_{ijmn}c^\dagger_{im\mu}c_{jn\mu}+K\sum_{ijj^\prime mn}\mathbf{S}_i \cdot c^\dagger_{jm\mu}\sigma_{\mu\nu}c_{j^\prime n\nu}\\
    &+J\sum_{<ij>}\mathbf{S}_i\cdot\mathbf{S}_j,
\end{split}
\label{eq1}
\end{equation}
where $c^\dagger_{im\mu}$ creates one hole with spin $\mu$ in orbital $m$ at site $i$ which is limited in two ligand $p$ orbitals, $\mathbf{S}_i=S\sum_{\mu\nu}d^\dagger_{i\mu}\bm{\sigma}_{\mu\nu}d_{i\nu}$ is the local spin in $d$ orbitals at site $i$, and $\bm{\sigma}_{\mu\nu}$ are the Pauli matrices.
The first term is the kinetic energy of itinerant carriers.
The second term describes the coupling between local spins and itinerant holes with $K\sim t_{pd}^2(\frac{1}{\Delta_{CT}}+\frac{1}{U_{dd}-\Delta_{CT}})$, and the third term is the AFM coupling between two neighboring spins in transition metals.
Here the local $S$ can vary from $1/2$ to $5/2$ depending on the electronic state of transition metal $d$ orbitals.
The exchange parameter $J$ between two local spins is determined by the energy gaining $\Delta E=2t_{pd}^4(\frac{1}{\Delta_{CT}^2U_{dd}}+\frac{2}{\Delta_{CT}^2(2\Delta_{CT}+U_{pp})})$ through superexchange processes.
(Note that in the above Hamiltonian we ignore the energy shift of the ground-state energy for simplicity.)

\begin{table} 
    \caption{Comparison of estimated charge transfer gap $\Delta_{CT}$, hybridization between $d_{x^2-y^2}$ and $p_{\parallelsum}$ orbitals $t_{pd}$, energy gaining from superexchange process $\Delta E$, N\'eel temperature $T_N$~\cite{Corey2014,Tsujimoto2007,Vaknin1989}, and the highest transition temperature $T_c$~\cite{Li2020,Balestrino2001,Azuma1992} under ambient pressure for different platforms. Due to the self-doping effect~\cite{Zhang2020,Botana2020,Karp2020,Lang2020}, LaNiO$_2$ is not AFM ordered but still has short-range correlation.}
    \begin{ruledtabular}
       \begin{tabular}{llll|lll}
          $d^{n}$&Mat.& $\Delta_{CT}$ & $t_{pd}$ & $\Delta E$  &$T_N$& $T_c$            \\\hline
$d^5$&SrFeO$_2$F  &$\sim$~5.0 &1.50 &$\sim$~0.11&$\sim$ 700K &?\\ \hline
$d^6$&SrFeO$_2$ &$\sim$~5.5 &1.45 &$\sim$ 0.075 &$\sim$ 470K&?\\\hline
$d^7$&LaCoO$_2$F  &$\sim$~5.5 &1.40 &$\sim$~0.066&? &?\\ \hline
$d^8$&LaNiO$_2$F &$\sim$ 5.0 & 1.30 &$\sim$ 0.061&?&?\\ \hline
 \multirow{2}{*}{$d^9$}&CaCuO$_2$  & $\sim$ 4.5        & 1.28      &  $\sim$ 0.075 & $\sim$ 500K &  $\sim$100K      \\
&LaNiO$_2$ & $\sim$ 6.0        & 1.30      &  $\sim$ 0.039   & -- & $\sim10$K
\end{tabular}
    \end{ruledtabular}
    \label{tab1}
\end{table}

Table~\ref{tab1} shows the comparison of $\Delta E$ for all different platforms.
For simplicity, all materials are assumed to have the same local repulsion of $U_{dd}=8$~eV, and additional repulsion in $p_{\parallelsum}$ orbital $U_{pp}=4$~eV~\cite{Ogata2008}.
Given the crude estimation of the charge-transfer gap, these values may deviate from the real case.
In addition, the intra-atomic repulsion for Fe and Co is usually smaller than 8~eV which can also lead to errors in our estimations.
Nevertheless, if the transition metals or electronic states($d^{n}$) are the same, the relative value between different materials should be robust, such as the enhancement of $\Delta E$ from SrFeO$_2$ to SrFeO$_2$F, and from LaNiO$_2$ to CaCuO$_2$.

In short, it's clear that these different materials all share the similar features in their electronic structure, namely low-energy states dominated by ligand $p$ orbitals, and orbitals other than $d_{x^2-y^2}$ in high energy due to strong Coulomb repulsion and decoupled from the mostly relevant $p_{\parallelsum}$ orbitals.
These two reasons guarantee the similar low-energy physics of hole carriers to the cuprates and nickelates.
The additional local spins in other $d^n$ (n$\neq$9) platform only influence the carriers through the spin channels.
Thus, the only necessary condition for the superconductivity in the cuprates analogs, whose relevant subspace is just $d_{x^2-y^2}$ orbital of transition metal and $p$ orbitals of ligand, probably is the single occupation of $d_{x^2-y^2}$ orbital in the parent compound.

The investigation on these sets of materials will be of significant importance since this may break our knowledge of $d^9$ electronic configuration in the prototype of square BX$_2$ plane that can host superconductivity such as CuO$_2$ and NiO$_2$ plane.
Platforms from $d^8$ to $d^5$ will provide different families of superconductors but all share similar low-energy physics which is helpful to understand the physics of HTSC.
Especially, with the systematic increase of the local magnetic moment from 1/2 to 5/2, the corresponding energy scale associated with the spin fluctuation will be lower.
This gradually makes the system evolving from an unable-to-use-perturbation case to one that can be treated perturbatively.
Thus, if superconductivity can be found in these materials, we definitely have a controlled approach to truly solve the long-standing puzzles of unconventional superconductivity.

From the consideration of better superconductors, the higher-energy $d$ orbitals of these elements other than Cu also open up rich tunability to improving superconducting temperature.
In the cuprates, the energy of Cu-$d$ orbitals is extremely low leading to a more stable $d^{10}\underline{L}$ state for most ligand elements except for O and F.
This causes no tunability in the choice of ligands to enhance the charge-transfer nature which is related to $T_c$~\cite{Weber2012} in the cuprates.
However, here, the energy of Fe (or Co, Mn) $d$ orbitals is higher, and thus other elements with high-energy $p$ orbitals, such as S/Se (or P/As) can also keep the single occupation of $d_{x^2-y^2}$ orbital in the parent compounds.
This ensures the possibility to tune the systems via ligand substitution to realize better superconductivity in these $d^1_{x^2-y^2}L$ types of materials.

We want to remind readers that these analyses based on electronic structure do not consider the complicated interaction between doped holes and phonons/magnons.
The remaining important concern is whether doped holes become localized due to the existence of spins in other $d$ orbitals or strong electron-phonon coupling (EPC)~\cite{Anisimov1992,Zaanen1994,Dobry1994,Bi1993} and thus the further superconductivity would be destroyed like that in La$_2$NiO$_4$~\cite{Cava1991}.
If the large spin ($S>\frac{1}{2}$) is not the decisive factor for the localization, other factors such as strong EPC can be probably avoided in the materials discussed here or via the tuning of some external conditions if we know the high-energy origination of strong EPC, such as the large charge-transfer gap~\cite{Zaanen1994}, or the presence of apical oxygen holes~\cite{Kuiper1991,Kuiper1998}.
Of course, these all need more experimental explorations.

These results could be easily verified in either hole-doped SrFeO$_2$ or SrFeO$_2$F via chemical or electrical doping, or topotactic fluorination of superconducting Nd$_{1-x}$Sr$_x$NiO$_2$.
To date, there is only electron-doped SrFeO$_2$ which shows good metallic behaviors~\cite{Katayama2019} while no experiments addressing the hole-doped SrFeO$_2$.
This is probably due to the unstable structure in chemical doping via K/Rb substitution.
Nevertheless, adding holes through gate voltage should be applicable in reality~\cite{Bollinger2011}.
If these materials are proven to be superconducting, we definitely open up a huge amount of tunability of the superconductivity in the cuprate analogs.
Achieving better superconductivity than that of the cuprates under higher temperature and ambient pressure also becomes promising, for example in Fe$^{3+}$/Co$^{3+}$-based materials.

\section{Summary}
\label{summary}
In summary, to explore the possible superconductors analogous to the cuprates, we investigate the electronic structure of materials with the same square BX$_2$ plane from $d^8$ to $d^5$ configuration in transition-metal ions.
We propose the possible candidate fluorinated nickelates LaNiO$_2$F to host superconductivity as the $d^8$ platform.
The extremely low energy of F $p$ orbitals excludes the interference from $d_{3z^2-r^2}$ orbital and apical F $p$ orbitals to the low-energy physics.
Corresponding electronic structures demonstrate good similarities between LaNiO$_2$F and the cuprates except for the additional upper Hubbard from $d_{3z^2-r^2}$ orbital.
The $d^7$ platform is allowed in Co$^{2+}$-based materials, such as LaCoO$_2$F.
Specially, the low-spin state $d_{x^2-y^2}^1d_{3z^2-r^2}^0t_{2g}^6$ in $d^7$ platform share the same feature with $d^9$, namely only one local spin in $d_{x^2-y^2}$ orbital.
The only difference is the empty $d_{3z^2-r^2}$ orbital which is in high energy and thus carries no weight to the low-energy physics.
This suggests the perfect similarity between LS $d^7$ and $d^9$ platform.
$d^6$ platform can be found in currently existing material SrFeO$_2$ which also shows similar electronic structures with the cuprates.
Last $d^5$ platform can be realized in SrFeO$_2$F which is already synthesized in reality.
As similar to LaNiO$_2$F, extremely low-energy $p$ orbital of F help Fe $d_{x^2-y^2}$ and O $p$ orbitals to be the only relevant subspace for hole carriers.
Overall, we find great similarities among all these materials in their low-energy states which are all dominated by the ligand X $p$ orbitals.
The strong charge-transfer nature of these materials suggests the doped holes should reside in ligand $p$ orbitals and transition metal $d_{x^2-y^2}$ orbitals as similar to the cuprates.
The irrelevant $d$ orbital other than $d_{x^2-y^2}$ are all in high energy as upper/lower Hubbard band due to strong intra-atomic Coulomb repulsion such that they barely influence the charge-related physics of low-energy carriers.
This leads to similar Hamiltonians to that of the cuprates but with a different number of local spins in other transition-metal $d$ orbitals than $d_{x^2-y^2}$ which can influence the low-energy carriers via spin correlation.
The role of these large spins in the transport of doped holes needs more investigation in the future.
Our results suggest the possible superconductivity in these materials despite the different electronic states of $d^n$.
The investigation of behaviors of hole carriers in these compounds may open up other platforms different from the $d^9$ configuration in high-$T_c$ superconductors.
The rich choice of transition metals and ligands also allows the tunability in the superconductivity to address the puzzles of unconventional superconductivity and to achieve better superconductors than the cuprates in the future.

\appendix
\section{Calculation details}
\label{Calculation_details}
In the calculations of electronic structure, we use LDA+$U$ method~\cite{Anisimov,Liechtenstein,Dudarev1998} with parameters $U-J= 8~\text{eV} - 0.8~\text{eV}=U_{\text{eff}}=7.2$~eV for all cases, implemented by Quantum Espresso~\cite{Giannozzi2020} with $7\times7\times7$ $\bm{k}$-mesh.
The pseudopotentials use PBE type~\cite{Prandini2018}, and wavefunction cutoff is 80~Ry and charge density cutoff is 800~Ry which are both kept fixed in all calculations.
All cases are assumed to be in $G$-type antiferromagnetic order.
We then extract the effective Hamiltonian $\mel{m}{H^{LDA+U}}{n}$ under the basis of Wannier function $\ket{n}$ using Wannier90~\cite{Marzari1997,Souza2001,Pizzi2020} and also the corresponding electronic structures in the Hartree-Fock (HF) approximation employed by LDA+$U$ calculations.

\section{LS $d^7$ state in Ni$^{3+}$}
\label{app2}
LS $d^7$ state with only one electron in $e_g$ orbitals is possible to realize a single occupation of $d_{x^2-y^2}$ orbital with $S=1/2$ like the cuprates which thus is expected to be a new platform to host superconductivity~\cite{Anisimov1999,Chaloupka2008}.
We have explored this issue on Co$^{2+}$ in Section~\ref{d7}.
Another candidate is LS Ni$^{3+}$, which has been proposed before that seems to be possible to realize via in-plane strain to break the degeneracy of $e_g$ orbitals through some heterostructures~\cite{Chaloupka2008,Hansmann2009,Disa2015}.
However, a true insulator state with singly occupied $d_{x^2-y^2}$ and fully empty $d_{3z^3-r^2}$ has not been synthesized in reality yet.

Here, we will discuss the realizability of LS $d^7$ state in Ni$^{3+}$ ion.
Starting with the most stable $d_{x^2-y^2}^1d_{3z^2-r^2}^1t_{2g}^6$ ($d^8$) state in Ni$^{2+}$, to make the additional hole residing in $d_{3z^2-r^2}$ orbital, the energy of $d_{x^2-y^2}^1d_{3z^2-r^2}^0t_{2g}^6$ needs to lower than that of $d_{x^2-y^2}^1d_{3z^2-r^2}^1t_{2g}^6\underline{L}$.
Due to the strong hybridization between $d_{x^2-y^2}$ and ligand $p_{\parallelsum}$ orbitals, the energy of the possible ground state contributed by $d_{x^2-y^2}^1d_{3z^2-r^2}^1t_{2g}^6\underline{L}$ will be further lowered via the formation of ZRS~\cite{Zhang}.
Thus, the energy of $d_{3z^2-r^2}$ needs to be extremely low (in hole picture) to realize the LS $d^7$ state in Ni$^{3+}$.

To investigate this question, let's start from LaNiO$_2$F with a stable $d^{8}$ state.
We first perform the LDA+$U$ calculation under AFM order.
The resulting electronic structure is shown in Figure~\ref{fig1}(b).
We then extract a many-body effective Hamiltonian $H^{\text{eff}}=\sum_{ijnm\mu}t_{ijnm\mu}c^\dagger_{in\mu}c_{im\mu}+\sum_{im1m2m3m4\mu\nu}U_{m1m2m3m4\mu\nu}c^\dagger_{m1\mu}c^\dagger_{m2\nu}c_{m3\nu}c_{m4\mu}$ of hole carriers in 10~eV scale within the subspace of Ni $d_{x^2-y^2},d_{3z^2-r^2}$ and three ligand $p_{\parallelsum}$ orbitals~\cite{Lang2020}.
$\mel{m1\mu,m2\nu}{U}{m3\nu,m4\mu}$ can be obtained via Slater integral~\cite{Slater,Liechtenstein} with the only two parameters $U_{\text{eff}},J_{\text{eff}}$.
Under the same HF approximation, $H^{\text{eff}}$ should have relation, $\mel{m}{H^{\text{eff}}}{n}=\mel{m}{H^{LDA+U}}{n}$.
Here $U_{\text{eff}}=7.82$~eV, $J_{\text{eff}}=1.14$~eV can well meet this requirement.

With this many-body Hamiltonian, the energy of each many-body state can thus be found.
The onsite energy in hole picture of each orbital is $\epsilon_{d_{x^2-y^2}}=-9.38$~eV, $\epsilon_{d_{3z^2-r^2}}=-9.03$~eV, $\epsilon_{\rm{O}-p_{\parallelsum}} = 2.81$~eV, $\epsilon_{\rm{F}-p_{\parallelsum}} = 6.32$~eV.
Therefore, the energy of $d_{x^2-y^2}^1d_{3z^2-r^2}^1t_{2g}^6\underline{L}$ state is $\epsilon_{d^8\underline{L}}=-8.34$~eV, while the energy of $d_{x^2-y^2}^1d_{3z^2-r^2}^0t_{2g}^6$ is $\epsilon_{\text{LS}-d^7}=-3.17$~eV.
(In fact, this even does not consider the anti-bonding combination of 4 O $p$ orbital around Ni atom which can further lower the energy of $d^8\underline{L}$ state to eV scale~\cite{Zhang}.)
Such a large energy difference, $\Delta \epsilon=\epsilon_{\text{LS}-d^7}-\epsilon_{d^8\underline{L}}$, of the order of 5-eV scale in fact indicates that it's very hard to realize this LS spin state in Ni$^{3+}$ via the splitting of crystal field.
Even if the energy of $d_{3z^2-r^2}$ is lower than that of $d_{x^2-y^2}$ for $1$~eV, this energy difference can only reduce to 3.7~eV.
This is understandable since to realize the double holes occupation of $d_{3z^2-r^2}$ orbital, one needs a very large charge-transfer gap to overcome the intra-atomic repulsion within $d$ orbitals while the CT gap here is not that large.
This result is consistent with our LDA+$U$ calculations in Section~\ref{d7} where we show that part of $d_{3z^2-r^2}$ band is below the fermi energy, and a full empty $d_{3z^2-r^2}$ orbital cannot be established from the consideration of the energy of O $p$ and Ni $d$ orbitals as predicted before~\cite{Anisimov1999}.
In addition, we note that the energy of $d_{xy}$ orbital is $\epsilon_{d_{xy}}=-8.53$~eV which leads to a high-spin $d_{x^2-y^2}^1d_{3z^2-r^2}^1d_{xy}^1d_{xz,yz}^4$ state with the energy $\epsilon_{\text{HS}-d^7}=-7.11$~eV that is also much lower than that of LS $d^7$ state because of the strong Hund's coupling in Ni.
Therefore, we expect that the desired electronic configuration $d_{x^2-y^2}^1d_{3z^2-r^2}^0t_{2g}^6$ is hard to realize in NiO$_2$-based materials.

\section{Possible electronic structure with HS-Co$^{3+}$}

\begin{figure}
	\begin{center}
		\includegraphics[width=\columnwidth]{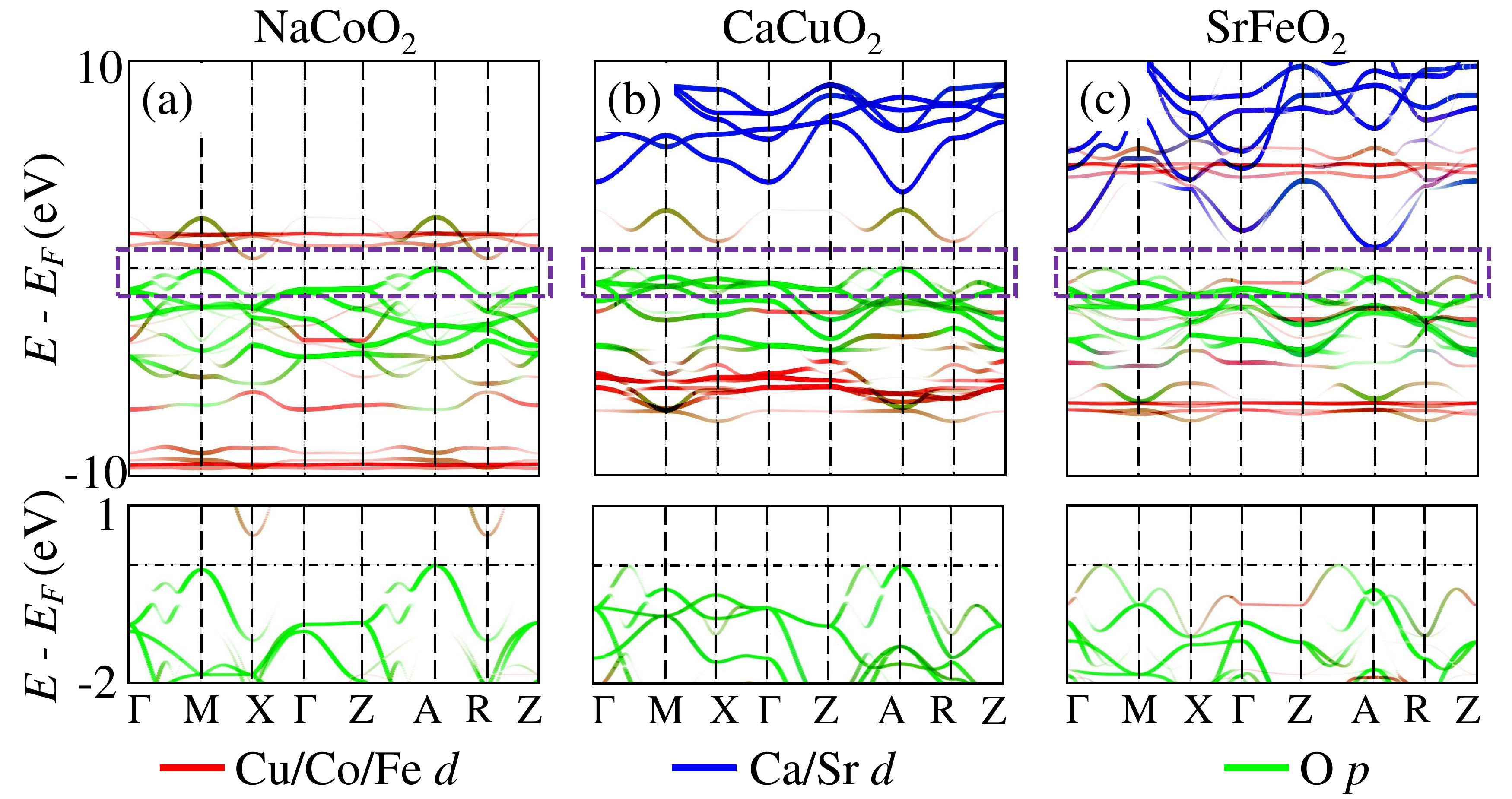}
	\caption{Comparison of unfolded electronic structure between (a) NaCoO$_2$, (b) CaCuO$_2$, and (c) SrFeO$_2$. Red, blue and green colors denote the weight of transition-metal $d$, Ca/Sr $d$, and O $p$ orbitals respectively. Low panels show the zoom-in band structures of each material near Fermi energy denoted by purple boxes in the upper panels. All under the same $G$-type AFM ordered state.}
	\label{d6bsCo}
	\end{center}
\end{figure}

Given the extremely low-energy $d$ orbitals of Co$^{3+}$ which in principle allows a strong magnetic correlation, it's valuable to explore the possible electronic structure with Co$^{3+}$.
For simplicity, let's consider NaCoO$_2$ materials as an example.
We first optimize the lattice structure using DFT+$U$~\cite{Dudarev1998} to ensure the correct orbital occupation for Co $d$ orbitals.
(This is implemented by Quantum Espresso~\cite{Giannozzi2020} using PBE pseudopotential~\cite{Prandini2018}.)
The resulting lattice constants are $a=b=3.815~\text{\AA}, c=3.406~\text{\AA}$ with the same space group as SrFeO$_2$.
Note that our goal here is not to find the stable structure, but rather to investigate its possibility to host strong magnetic correlation and further superconductivity based on its electronic structure.

Figure~\ref{d6bsCo} shows the resulting band structure of NaCoO$_2$ under AFM ordered phase with the same parameter $U_{\text{eff}}=7.2$~eV and its comparison to CaCuO$_2$ and SrFeO$_2$.
All of them show well-defined insulating states, and the band gap in NaCoO$_2$ is the smallest which is only about 0.6~eV.
There are four upper Hubbard bands corresponding to the high-spin $d^6$ state.
The only doubly occupied orbital is $d_{3z^2-r^2}$ orbital similar to that of SrFeO$_2$ as we expect.

Compared to SrFeO$_2$, the position of the upper Hubbard band is much lower and more close to that of the cuprates as a result of low-energy $d$ orbitals in Co$^{3+}$.
This leads to a small charge-transfer gap similar to the cuprates.
Specially, the resulting coupling between Co $d_{x^2-y^2}$ and O $p_{\parallelsum}$ orbitals is $t_{pd}\approx1.60$eV which is larger than the $t_{pd}$ in CaCuO$_2$ and SrFeO$_2$.
One thus expects a very strong magnetic correlation in NaCoO$_2$.
Assuming the CT gap $\Delta_{CT}\approx4.5$~eV, $U_{dd}=8~\text{eV}, U_{pp}=4\text{eV}$, and the fourth order perturbation is still applicable, the resulting energy scale of AFM correlation is $\Delta E\approx0.181$~eV.
This indicates the N\'eel temperature of AFM order will be very high.
If the superconducting state can be hosted in hole-doped samples, the resulting superconducting phase probably can also sustain to a high temperature under such a large energy scale.

\bibliography{MainTex.bib}
\end{document}